\documentclass[12pt]{iopart}

\usepackage[dvips]{graphicx}
\usepackage{epsfig}
\usepackage{color}

\def\jpb{{\em J. Phys. B: At. Mol. Opt. Phys.}~}
\def\pra{{\em Phys. Rev. A}~}

\def\prl{{\em Phys. Rev. Lett.}~}

\def\jetp{{\em Sov. Phys. JETP}~}

\newcommand{\beq}{\begin{equation}}
\newcommand{\eeq}{\end{equation}}
\newcommand{\vecp}{{\bf p}}
\newcommand{\vecv}{{\bf v}}
\newcommand{\vecA}{{\bf A}}

\newcommand{\vecr}{{\bf r}}

\newcommand{\cale}{{\cal E}}

\begin{document}

\title{Laser assisted decay of quasistationary states}

\author{H~M~Casta\~neda~Cort\'es$^1$, S~V~Popruzhenko$^2$, D~Bauer$^3$ and A~P\'alffy$^1$}
\address{$^1$Max-Planck-Institut f\"ur Kernphysik, Saupfercheckweg 1, 69117 Heidelberg, Germany}
\address{$^2$Moscow Engineering Physics Institute,  National Research Nuclear University, Kashirskoe Shosse 31, 115409 Moscow, Russia}
\address{$^3$Institut f\"ur Physik, Universit\"at Rostock, 18051 Rostock, Germany}

\ead{Hector.Castaneda@mpi-hd.mpg.de}
\date{today}

\begin{abstract}
The effects of intense electromagnetic fields on the decay of quasistationary states are investigated theoretically.
We focus on the parameter regime of strong laser fields and nonlinear effects where an essentially nonperturbative description is required. Our approach is based on the imaginary time method previously introduced in the theory of strong-field ionization. Spectra and total decay rates are presented for a test case and the results are compared with exact numerical calculations. 
The potential of this method is confirmed by good quantitative agreement with the numerical results.

\end{abstract}

\maketitle

\section{Introduction}
Since the 1960s when lasers became a worldwide-used laboratory equipment, experimental and theoretical studies of nonlinear phenomena emerging in the interaction of intense electromagnetic fields with matter have seen continuous progress.
One of the key phenomena in the physics of strong-field laser-matter interaction is nonlinear ionization.
Although observed for the first time in 1965 \cite{delone65}, it continues to attract attention as a fundamental process that  encodes pivotal information on quantum dynamics in  the presence of intense electromagnetic fields.
Another reason why nonlinear ionization remains in the scope of theoretical and experimental research is that it is a building block for numerous complex laser-induced and laser-assisted phenomena, including the generation of high order harmonics, nonsequential multiple ionization and a variety of effects in laser plasmas.

Typically, the quantum system subject to an intense laser pulse is assumed to be stable in the absence of the field.
In this sense, ionization is viewed as a laser-induced effect.
This indeed holds for a variety of problems related to the interaction of laser pulses with atomic and molecular gases.
Within this common picture, ionization differs qualitatively from  laser-assisted phenomena, i.e.,  effects that also take place without the presence of an external field but can yet be essentially modified by it.
Bremsstrahlung in external electromagnetic fields is such an example  \cite{bunkin66,fedorov-book}. Ionization can also be considered a laser-assisted effect under specific conditions, such as
when the system is prepared in an excited state with a positive total energy  (e.g., autoionizing states) or is under the action of a constant or slowly varying electric field and is thus free to decay by itself. In this case, an additional external field can influence the decay by modifying the electron spectra and the total probability.  One may thus interpret such a process as being a laser-assisted decay (or, equivalently, ionization) from a quasistationary (QS) state.

A QS state in the presence of an AC field can be in many cases formally treated as a true bound state affected by two fields out of which one is static or quasistatic.
From this point of view, laser-assisted decay is nothing but ionization in a two-color field.
Nonlinear ionization and other related processes in the field of two or more superimposed intense laser pulses of different frequencies have been considered in theoretical studies.
In particular, the interaction of laser radiation with negative ions in the presence of a static electric field was treated in detail in Ref.~\cite{manakov-rev00}, where the authors used the QS quasienergy states formalism, which combines the QS state and Floquet approaches. In addition, a number of theoretical works have considered two-color strong-field ionization. However, there is an essential difference in the statement of the problem that limits the possible application of the aforementioned results to the description of laser-assisted tunneling of a QS state: in the case of two-color laser-induced ionization, if one of the two fields is turned off, the photoelectron spectrum remains an above-threshold-ionization (ATI) spectrum that consists of several ATI maxima separated by the photon energy. This holds for any ac field, irrespective of its frequency. If the second field is static (as considered in Ref.~\cite{manakov-rev00}) then there is no spectrum along the direction of the electric field vector, since the respective momentum component is not conserved.
In this case, the only observable of the problem is the rate of ionization, or the width of the QS state.
When, however a true QS state is considered, in the absence of the laser field the spectrum has the form of a single narrow line centered at the real part of the QS state energy. None of the two aforementioned cases possesses this limit.

The aim of this work is to develop a theoretical description of laser-assisted decay of  QS states. For laser-induced ionization, a simple and simultaneously very efficient nonperturbative theory was proposed by Keldysh \cite{keldysh}. This approach, meanwhile known in its different realizations as the Keldysh-Faisal-Reiss model \cite{faisal,reiss}  or the Strong-Field Approximation (SFA) \cite{reiss} (for the  present status of the SFA and its implementations see Refs. \cite{popov-rev,milos-rev}), is a well-established  tool in strong-field physics. The essence of the SFA is that it approximates the electron continuum by the Volkov states, which are exact solutions of the Schr\"odinger (Dirac) equation in the presence of a plane electromagnetic wave \cite{volkov,dau4}. The effect of the binding potential on the electron continuum is simultaneously neglected.
This key assumption of the SFA is intuitively natural and provides a way to a simple description of strong-field laser-atom interactions that in many cases is fully analytical.

Here we adopt
the general idea of SFA and modify it for the description of  laser-assisted decay. For this task, we use the  formulation of the SFA in terms of complex classical trajectories known as the Imaginary Time Method (ITM). This method was introduced in the early days of strong-field physics \cite{popov67,kuznetsov} to give a physically transparent formulation of the Keldysh method. In addition, the ITM  also provides an efficient way to consider significant effects that the Keldysh model misses in its original formulation\footnote{In particular, the ITM was used to account for the effect of long-range Coulomb interaction on the ionization dynamics. It is currently proven in experiments and numerical simulations that the Coulomb interaction strongly affects nonlinear ionization. Coulomb corrections to the ionization amplitude derived within the ITM provide a quantitative description of total ionization rates \cite{popov67,poprz-prl08} and photoelectron momentum distributions \cite{bauer08,science11,tianmin}.
The ITM and its applications were recently reviewed in Ref. \cite{popov-itm}.}.
% end footnote
The idea of using the ITM for the description of the tunneling of free particles through a potential barrier in the presence of an oscillating electric field was first proposed by Ivlev and Melnikov \cite{ivlev85}.
 Here, using the ITM,  we develop a more general and accessible theoretical description of the  laser-assisted decay of QS in the presence of intense electromagnetic fields. Our approach recovers specific results of Ref. \cite{ivlev85}  and goes beyond by providing not only qualitative but also quantitative tunneling probabilities that agree with exact numerical calculations. 
In the form presented here, restricted to the nonrelativistic limit, our method should be equally applicable for descriptions of different physical systems under standard semiclassical conditions. 

The general question about how AC electric fields can influence the decay of QS, i.e, tunneling, has  been theoretically addressed in other fields. Most of the work we are aware of relates to the tunneling between superconductor films or micro- and  nanocontacts affected by a microwave, infrared or optical laser field.
For such systems the effect is commonly referred to as photon-assisted transport (see \cite{platero04} and references therein).
Typically, in experiments on photon-assisted transport,  low and moderate intensity laser fields  are applied so that, although the coupling is nonlinear and multiphoton processes are present or even dominate, the characteristic number of absorbed or emitted photons is not very large. For such a moderately nonlinear regime of interaction, the Floquet method is known to be an efficient approach, and is routinely used \cite{platero04}.
However, when a large number of photons is involved, the Floquet approach becomes less efficient and more cumbersome numerically. On the other hand, the ITM is particularly relevant in the semiclassical domain where a typical number of photons is at least on the order of 10 and can easily reach hundreds and even thousands.
Thus, the approach we develop here can be viewed as a complementary one with respect to the Floquet method
and may provide a new efficient computational tool in theoretical studies of photon-assisted transport in the highly nonlinear coupling regime.

Another example of laser-assisted decay is beta-decay in the presence of electromagnetic fields \cite{nikishov-beta}. The decay of elementary particles can be modeled within the picture of QS states only on a qualitative level. In the original work of Nikishov and Ritus \cite{nikishov-beta}, a more rigorous approach was used, based on the low-energy limit of the weak interaction theory modified to account for the effect of the electromagnetic field.
To this end,  the muon and the electron plane waves were replaced by Volkov waves. 
As will be shown later in this work, our approach reproduces an important qualitative conclusion of earlier studies  \cite{nikishov-beta,scully83,scully84,scully84a}, namely that depending on the parameters, two different regimes of decay are realized (i) when the spectrum is strongly affected without a modification of the total decay rate and (ii) when the rate of decay is also affected. 
These two regimes are also referred to as  ``exclusive'' and ``inclusive'', respectively.
Although Refs. \cite{nikishov-beta,scully83,scully84,scully84a} consider the influence of laser fields on the decay of elementary particles, the qualitative classification of the interaction regimes holds for a variety of laser-assisted processes.

The physical difference between the two regimes becomes clear if we notice that for a strong modification of the spectrum no high-field intensity is actually  needed, but only a large quiver (ponderomotive) energy.
The latter can be achieved at low laser frequencies.
A relatively weak but low-frequency laser field strongly affects the kinematics of the charged particle, accelerating or decelerating it after decay.
This changes the final energy at the detector with almost no effect on the total decay probability.
In contrast, in the ``inclusive'' regime, the particle's dynamics on the short-time scale corresponding to the subbarrier motion is also influenced, modifying the total probability.
This indeed requires high laser field strengths.
The particular expression of the critical field that delimits the ``inclusive'' regime depends on the investigated system. 
For beta-decay such fields lie far beyond present experimental capabilities (see, e.g., Ref. \cite{scully84a} and references therein) and are probably in principle unachievable \cite{fedotov10}, whereas for atomic and solid state systems discussed in this work, both regimes are accessible.

This paper is organized as follows.
In Section \ref{SFA} we introduce the SFA and adapt it to describe the laser-assisted decay
of QS states. The reformulation of SFA in terms of the ITM is used to derive the transition amplitude. With the help of the latter, we determine in section \ref{sqbarrier} the ATI-like spectra of a model one-dimensional (1D) problem where the QS state can tunnel through  a rectangular barrier. The spectra are analyzed and compared with the results of the exact numerical solution of the time-dependent  Schr\"odinger equation (TDSE). We  conclude with a short summary and outlook. Atomic units $\hbar=m_ e=| e|=1$ are used throughout the paper.

%%%%%%%%%%%%%%%%%%%%%%%%%%%%%%%%%%%%%%%%%%%%%%%%%%%%%%%%%%%%%%%%%%%%%%%%
\section{Basic equations\label{SFA}}
%%%%%%%%%%%%%%%%%%%%%%%%%%%%%%%%%%%%%%%%%%%%%%%%%%%%%%%%%%%%%%%%%%%%%%%%

\subsection{Strong-field approximation (SFA) matrix element for quasistationary (QS) states \label{SFA-various}}
%%%%%%%%%%%%%%%%%%%%%%%%%%%%%%%%%%%%%%%%%%%%%%
We start from a short summary of the SFA that describes ionization from true bound states.
Within the SFA, the transition amplitude between an atomic bound state $|\Psi_0\rangle$ of binding energy $E_0\equiv-I$ and a continuum state $|\Psi_{\vecp}\rangle$ with an asymptotic momentum $\vecp$ is given by
\beq
M_{\rm SFA}(\vecp)=-i\int\limits_{-\infty}^{+\infty}\langle\Psi_{\vecp}\vert\hat V(t)\vert\Psi_0\rangle dt\, ,
\label{MSFA}
\eeq
where the final state is approximated by the Volkov function,
\beq
\Psi_{\vecp}(\vecr,t)=\frac{1}{(2\pi)^{3/2}}\exp\left\{i\vecv_{\vecp}(t)\cdot\vecr-i\int\limits_{-\infty}^t
\varepsilon_{\vecp}(t^{\prime})dt^{\prime}\right\}
\label{Volkov}
\eeq
and
$$
\varepsilon_{\vecp}(t)=\vecv_{\vecp}^2(t)/2,~~{\mathrm{with}}~~~\vecv_{\vecp}(t)=\vecp+\vecA(t)/c
\nonumber
$$
are the electron time-dependent kinetic energy  and velocity in the electromagnetic field,respectively, described by the vector potential $\vecA(t)$. $\hat V(t)$  is the interaction operator of the electron with the field of the electromagnetic wave and $c$ the speed of light.
In the dipole approximation, the electric field ${\cale}(t)=-\partial_t\vecA/c$ and the vector potential depend only on time.
It is convenient to simplify the notation by using the field-induced momentum rather than the vector potential, $\vecp_F(t)=\vecA(t)/c$.

Amplitude (\ref{MSFA}) is relevant under the semiclassical conditions:
\beq
K_0=\frac{I}{\omega}\gg 1\, ,~~~~F=\frac{{\cale}_0}{{\cale}_{\rm ch}}\ll 1\, ,
\label{K0}
\eeq
where $\cale_0$ is the electric field amplitude, ${\cale}_{\rm ch}=(2I)^{3/2}$ is the characteristic atomic field (for the ground state of hydrogen ${\cale}_{\rm ch}=m_e^2 e^5/\hbar^4=5.14\times 10^9$V/cm)  and $\omega$ is the laser frequency.
The first strong inequality in (\ref{K0}) shows that the minimal number of photons required for ionization is large, hence the coupling is essentially nonlinear. The  second inequality guarantees that the spatial scale on which the ionization amplitude forms is large in comparison with the atomic size (see a detailed discussion in Refs.\cite{popov-rev,bauer08}).
 Another frequently used dimensionless combination known as the Reiss parameter \cite{reiss}  is proportional to the ratio of the ponderomotive energy $U_P=\langle\vecp_F^2(t)\rangle_T/2$ (where $T$ is the optical period)  to the photon energy.
For the linearly polarized monochromatic field this reads
\beq
z_F=\frac{4U_P}{\omega}=\frac{\cale_0^2}{\omega^3}=8F^2K_0^3\, .
\label{zF}
\eeq
For ionization of atoms and positive ions by intense infrared and optical lasers, the conditions (\ref{K0}) are usually well satisfied and  $z_F\gg 1$.
The integrand in (\ref{MSFA}) is then a rapidly oscillating function of time.
This allows us to evaluate the integral by the saddle-point method, so that the amplitude can be written as a sum of contributions from all relevant stationary points $t_0(p)$,
\beq
M_{\rm SFA}(\vecp)=\sqrt{-2\pi i}\sum_{\alpha}{\cal P}(\vecp,t_{0\alpha})\frac{\exp\left(-iS_0(\vecp,t_{0\alpha})\right)}
{\sqrt{\partial^2_tS_0(\vecp,t_{0\alpha})}}\, ,
\label{MSFAsp}
\eeq
where $S$ is the classical action
\beq
S_0(\vecp,t)=\int\limits_t^{+\infty}\left\{\vecv_{\vecp}^2(t^{\prime})/2+I\right\}dt^{\prime}
\label{S0}
\eeq
and the pre-exponential factor ${\cal P}$ is the spatial matrix element of the interaction operator $\hat{V}$. 
The saddle-point equation is of the form
\beq
\partial_tS_0(\vecp,t_0)=\vecv_{\vecp}^2(t_0)/2+I=0\, ,
\label{speq}
\eeq
which shows that a saddle point is always complex for $I>0$.
The differential ionization rate is given by the squared modulus of (\ref{MSFAsp}).

The fact that all roots of Eq.~(\ref{speq}) are complex reflects the quantum nature of ionization.
Consequently, phase $S_0(\vecp,t_0)$ in (\ref{MSFAsp}) is a complex quantity with a numerically large and negative imaginary part, hence under conditions (\ref{K0}) the ionization probability appears to be a highly nonlinear function of the laser field strength.
The form of this nonlinear dependence is quantified by the value of the Keldysh parameter \cite{keldysh}
\beq
\gamma=\frac{\sqrt{2I}\omega}{{\cale}_0}\equiv\frac{1}{2K_0F}
\label{gamma}
\eeq
which is the ratio of the characteristic atomic momentum $\kappa_0=\sqrt{2I}$ to the field momentum $p_F=\cale_0/\omega$.
Since the final state is approximated by the plane wave (\ref{Volkov}), the prefactor ${\cal P}$ in (\ref{MSFAsp}) can be expressed via the Fourier transform of the bound state atomic wave function $\phi_0(\vecr)$.
For practical calculations, this means that the prefactor is a weak function of the final momentum and the field and atomic parameters compared with the highly nonlinear exponential, so that one may safely replace it by a constant as long as photoelectron momentum distributions and not the total ionization rates,  are considered.
Moreover, in the above formulation of the SFA, the so-called  plain SFA, the expression of the prefactor is anyway incorrect, except for the case of ionization from a short-range well. The simplest form of the SFA transition amplitude is therefore given by Eq.~(\ref{MSFAsp}) with ${\cal P}={\rm const}$.
On the qualitative level this rough approximation is in many cases sufficient. Strictly speaking, the SFA provides a quantitatively correct description of nonlinear ionization only for the exceptional case of a particle bound by a zero-range potential. For short-range potentials, it is still a good approximation if the interaction operator ${\hat V}(t)$ is taken in the length gauge \cite{bauer05}. For atoms, where the electron-core interaction potential always presents a long-range Coulomb tail, the SFA prefactor ${\cal P}$ is essentially wrong in any gauge. In this case, to calculate it correctly and bring the SFA back to the quantitative level of description, the technique of Coulomb corrections was developed. For further details  we refer the reader to Refs.~\cite{poprz-prl08,bauer08,smirnova08}.

\begin{figure}
\includegraphics[width=0.6\linewidth]{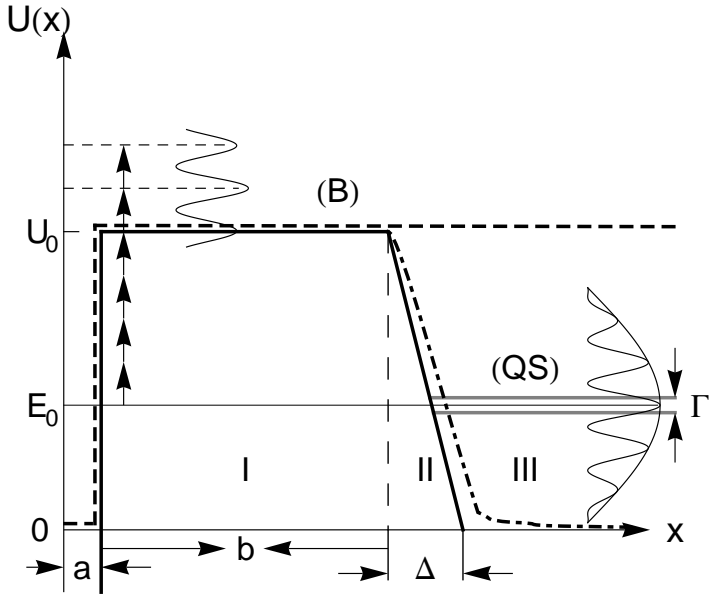}
\caption{Sketch of 1D short range potential $U(x)$ including the potential well ($0\le x \le a$) and rectangular barrier (I). A smooth boundary to the rectangular barrier, including the triangular barrier of width $\Delta$ (II), is shown by the dashed-dotted line. Qualitative photoelectron spectra for strong-field ionization from a true bound state (B) and a QS state (QS) are presented. Dashed line: the infinite barrier used to obtain numerically with the help of the Schr\"odinger equation the ground-state wavefunction of the particle inside the well. For such a barrier with $b=\infty$, only the ATI spectrum (B) is present. See Section \ref{comparison} for further details. }
\label{fig1}
\end{figure}

Let us now turn to the case when not a bound but a QS state is subject to an intense laser pulse.
One would then expect the appearance of an ATI-like photoelectron spectrum with the significant difference that now the initial state energy $E_0\equiv {p_0}^2/2>0$ so that there is no gap between the initial state and the continuum.
As a result, laser photons can also be emitted, not only absorbed, and the net number of absorbed photons can also be zero.
Figure \ref{fig1} sketches this qualitative difference between photoelectron spectra for stationary and QS states.
Nevertheless, in strong fields we expect that the typical number of photons involved in the interaction is anyway large, hence SFA-like approaches should be suitable also for the description of ionization from QS states.
With this assumption, we can introduce the amplitude of  laser-assisted decay replacing the bound state wave function $|\Psi_0\rangle$ in (\ref{MSFA}) by  the QS state (Gamow's wave function)
\[
\Psi_0(\vecr,t)=\phi_0(\vecr)e^{-iEt},~~~~~E\equiv E_0+iE_0^{\prime}=E_0-i\Gamma/2=\frac{{p_0}^2}{2}-ip_0p_0^{\prime}\, .
\nonumber
\]
Here $E_0$ is the real part of the complex energy and $\Gamma$ is the width that determines the decay rate.
Following a common width limitation  in the theory of QS states, we consider $\Gamma\ll E_0$.
If an even stronger limitation is satisfied and the width is small compared with all other characteristic frequencies of the problem, we may disregard the factor $\exp(-\Gamma t/2)$ in the integrand.
Then the SFA ionization amplitude differs from the one for the true bound state in the spatial wave function of the initial state $\phi_0(\vecr)$ and by the fact that the initial state energy $E_0$ lies in the continuum.

This straightforward application of the SFA leads, however, to some difficulties, namely:
\begin{enumerate}
{\item The spatial matrix element is divergent due to the exponential divergence of the spatial wave function of the QS state,
\[
\phi_0(\vecr)\sim e^{ip_0r+p_0^{\prime}r}\to\infty,~~~~r\to\infty,~~~~p_0^{\prime}\approx
\Gamma/2p_0\, .
\nonumber
\]
This asymptotic exponential divergence at large distances well known in the theory of QS \cite{bzp,peierls}}, originates from the approximate treatment of the decaying state as a stationary state and was noted in the pioneering work of Gamow \cite{gamow28}.
{\item Even if the phase is large and the saddle point method is applicable, the saddle points are real (since $E_0>0$ and we omit $\Gamma\ll E_0$ in the saddle-point equation) and the stationary phase is also real. The field parameters then enter the tunneling probability only via the pre-exponential factor. In other words, the transition amplitude does not demonstrate the nonlinear dependence on the laser field strength and frequency typical for strong field phenomena. This means that even in a very weak field the probability of detecting the outgoing particle with an energy considerably different from $E_0$ is not small, i.e. the field-free tunneling exponent does not emerge in the limit $\mathcal{E}_0\to 0$. Obviously, such a conclusion cannot be correct}.
\end{enumerate}

Although the exponential divergence of QS state wavefunctions is itself not surprising and follows from the definition of a QS state, for the calculation of norms and matrix elements containing these divergent factors a regularization method is needed.
Such a method was first proposed by Zeldovich \cite{zeldovich}.
However, for our purposes we will not use any regularization but apply instead another method for reconstruction of the correct prefactor, as explained in the following.

The origin of the second  difficulty becomes clear if we consider the structure of the continuum for the system shown in Figure \ref{fig1}.
For simplicity, in the following we refer to a 1D system.
At large distances the eigenfunctions are superpositions of the incoming and the outgoing plane waves:
$$
\phi_p(x)= e^{-ipx}+F(p)e^{ipx},~~~~~x\to\infty,~~~~~E=p^2/2
\nonumber
$$
while inside the well, $x\le a$,
$$
\phi_p(x)\sim A(p)\sin\left(\int\limits_0^xv(p,x^{\prime})dx^{\prime}\right)\, ,~~~~~v(p,x)=\sqrt{p^2-2U(x)}\, .
\nonumber
$$
For the momenta in the narrow vicinity of the QS state, $p\simeq p_0$, the absolute value of the coefficient $A(p)$ depends strongly on $p$ having a sharp maximum at $p=p_0$.
For eigenstates whose energy is sufficiently different from $E_0$,
\beq
\vert E-E_0\vert\gg\Gamma\, ,~~~~~0<E<U_0\, ,
\label{k-k0}
\eeq
the wave function inside the well is exponentially small,
\beq
A(p)\simeq e^{iW_0(p)}\, ,~~~~~W_0(p)=\int\limits_{a(p)}^{b(p)}v(p,x)dx\, .
\label{A}
\eeq
Here $a(p)$ and $b(p)$ are the turning points of classical motion, hence the action $W_0$ taken over the classically forbidden region is a purely imaginary value and $iW_0<0$.
The coefficient $A(p)$ is the semiclassical probability amplitude (calculated with exponential accuracy) of the particle to tunnel through the barrier formed by the potential $U(x)$.
Thus, for continuum states satisfying (\ref{k-k0}) the correct spatial matrix element should contain an exponentially small factor (\ref{A}), whereas the SFA matrix element calculated with a plane wave final state function does not present this feature.
We come to the conclusion that the problem with the SFA applied to QS states is that its plane wave final state Volkov function differs from the correct continuum wave function exponentially, exactly in that part of the position space that contributes most to the spatial matrix element.

Taking this into account we may formulate how one should modify the SFA matrix element to make it appropriate also for the description of LAT from QS states; the spatial matrix element should be replaced according to
\beq
\sqrt{-2\pi i}\,\langle \phi_p\vert V(x,t)\vert\phi_0\rangle/\sqrt{\partial^2_tS_0(p,t_{0\alpha})}\to A(v_p){\cal P}(v_p,t)\, ,
\label{new-spatial}
\eeq
where $A(v_p)$ is given by (\ref{A}), $v_p$ is the instant velocity $v_p(t)=p+p_F(t)$  and the new prefactor ${\cal P}(v_p,t)$ will be defined below.
Then, instead of (\ref{MSFAsp}) we obtain
\beq
M_{\rm SFA}(p)=\sum_{\alpha}{\cal P}
(v_p,t_{0\alpha})\exp\left[iW_0(v_p)-iS_0(p,t_{0\alpha})\right]
\, .
\label{MSFA-QS}
\eeq

The action $W_0(v_p)$ is  complex and describes the field-free tunneling through the potential barrier $U(x)$.
The action $S_0(p,t_0)$ is real (just as the saddle point $t_0$)  and therefore accounts for the effect of the laser field on the particle after the tunneling.
Thus, in this approximation the laser field  only changes the particle's energy on its way from the tunneling exit to a detector.
In other words, $S_0$ accounts for the kinematic effect of the laser field that redistributes the particles in the energy space not affecting the total decay probability.
 This corresponds to the ``exclusive''  regime of interaction \cite{nikishov-beta,scully83,scully84,scully84a}.
At first sight, the tunneling exponent is affected by the laser field via the fact that the field-free action $W_0$ is now taken at the instant velocity $v_p$ at the saddle point.
However, according to the saddle point equation (\ref{speq}) $v_p(t_0)=p_0$, so that $W_0(v_p)=W_0(p_0)$ and the laser field dependence vanishes.

We can now translate this formal description of the matrix element (\ref{MSFA-QS}) into a simple physical picture which would allow us to determine the correct prefactor ${\cal P}(p,t)$.
The particle tunnels through the potential barrier the same way as it would without the laser field.
At some  time instant $t_0$, it emerges in the classically allowed domain having the initial velocity $v(t_0)=p_0$ and starts its motion in the laser field\footnote{Here we assume that the potential well is a short-range one.}.
Then its final energy is given by
\beq
E=p^2/2=(p_0-p_F(t_0))^2/2,
\label{Efinal}
\eeq
so that each initial time $t_0$ corresponds to a certain final energy.
The inverse function $t_0(E)$ is not necessarily single-valued.
In the linearly polarized monochromatic field with
\beq
\cale(t)=\cale_0\cos\omega t\, ,~~~~~p_F(t)=-p_F\sin\omega t\, ,~~~p_F={\cale_0/\omega}
\label{et}
\eeq
we have $E_{\rm max}=(p_0+p_F)^2/2$ and $E_{\rm min}=(p_0-p_F)^2/2$ or zero; the spectrum consists of  ATI-like peaks between the classical  boundaries (CB) $E_{\rm min}\le E\le E_{\rm max}$.
The magnitudes of the ATI-like peaks vary slowly with the energy via the prefactor in Eq.~(\ref{MSFA-QS}).
Outside the CB, the spectrum vanishes abruptly.

Within the picture described by Eq.~(\ref{MSFA-QS}), penetration of the particle through the well and its subsequent evolution are independent.
Then the probability to tunnel out during the time interval $dt_0$ is given by
\beq
dw(t_0)=R_0dt_0=R_0\bigg\vert\frac{dt_0}{dp}\bigg\vert dp\, ,
\label{dw-new}
\eeq
where $R_0={\cal P}_0^2\exp(-2{\rm Im}W_0)$ is the field-free tunneling rate and the derivative $dp/dt_0=-\dot p_F(t)$ is calculated from (\ref{Efinal}). 
Note that we assume here that the tunneling probability can be written as a continuous function of the particle asymptotic momentum $p$. In reality, the spectrum consists of a discrete comb of peaks corresponding to absorption or emission of an integer number of laser photons. Our assumption thus implies that the characteristic number of peaks $L$ in the spectrum is large, $L\gg 1$.
According to Eq.~(\ref{Efinal}), there are two limiting cases defined by the ratio between the field-free electron asymptotic momentum $p_0$ and the field momentum $p_F$.
If $p_F\ll p_0$ then the classical boundaries of the spectrum are approximately $E_0\pm p_0p_F$ so that the number of peaks is $L\simeq p_0p_F/\omega$.
For stronger fields where $p_F\gg p_0$ the energy scale is determined by the ponderomotive energy and the number of peaks is of the order of the Reiss parameter (\ref{zF}).
As will be discussed in the next section, for field values for which $L$ is not much greater than unity, the approximation of the continuous spectrum contradicts  energy conservation requirements. As a result, the total rate can only be calculated with some numerical error.

The distribution (\ref{dw-new}) is divergent at the CBs $p=p_{\rm max/min}$ where we have $dp/dt_0=0$.
Such an integrable divergence near the CBs is typical for SFA-based descriptions,  occurring at the final momenta for which the saddle-point method does not work due to cancellation of the second derivative of the action.
This does not affect the total probability but renders the momentum distribution incorrect in the vicinity of CBs.
To avoid this problem,  the term that is proportional to the third derivative of the action has to be accounted for in the phase decomposition near the saddle point.
The divergence is then replaced by a local maximum of the spectrum at the CB \cite{gor99} with a relative height of the order of $z_F^{1/3}$, where $z_F$ is the Reiss parameter (\ref{zF}).
In the simplest form, this regularization procedure reduces to the replacement
\beq
\bigg\vert\frac{dt_0}{dp}\bigg\vert\to\frac{1}{\sqrt{(dp/dt_0)^2+\beta^2}}\, ,~~~~\beta=\cale_0^{1/3}\omega\, .
\label{regular}
\eeq
Next, we take into account that usually there is more than one solution to the saddle-point equation, so that several $t_{0\alpha}$, $\alpha=1,2,...$ correspond to the same final energy.
This leads to a coherent sum over all saddle-point solutions.
The momentum distribution takes then the form:
\beq
dw(p)=\left\vert M(p)\right\vert^2dp\, ,~
M(p)=\sum_{\alpha}\frac{{\cal P}_0\exp\left[iW_0(v_p)-iS_0(p,t_{0\alpha})\right]}{\sqrt{dp/dt_0+i\beta}}\bigg\vert_{t=t_{0\alpha}}\, .
\label{SFA-correct}
\eeq
 Clearly, this result misses two effects:
(i) the influence of the laser field on the subbarrier motion is not accounted for and (ii) in the classically allowed domain, the effect of the potential $U(x)$ is disregarded.
The former effect becomes more and more significant when the laser intensity grows, whereas the latter is particularly important for potentials with a long-range tail, e.g. for the Coulomb potential.
In the next subsection, we reformulate the amplitude (\ref{SFA-correct}) using the ITM and show that this new formulation provides a straightforward way to account for the two missing effects.

\subsection{Imaginary time method for quasistationary (QS) states}
%%%%%%%%%%%%%%%%%%%%%%%%%%%%%%%%%%%%%%%%%%%%%%%%%%%

The amplitude in Eq.~(\ref{SFA-correct}) can be equivalently reformulated in terms of classical complex trajectories.
According to the ITM \cite{popov-itm}, a trajectory $x_0(t)$ can be found along which the particle starts its motion at the complex time instant $t=t_s$ inside the well, $x(t_s)=0$, having the energy $E=v^2(t_s)/2=E_0$ and arrives at $x=b$ when $t=t_0$.
Here $b=b(E_0)$ is the outer classical turning point, $U(b)=E_0$.
The trajectory satisfies the Newton equation
\beq
\ddot x=\dot v=-{\partial U}/{\partial x}\, .
\label{NE0}
\eeq
The exit point $x=b$ is separated from the well by the classically forbidden region; hence, the solution of Eq.~(\ref{NE0}) satisfying the assigned initial conditions only exists in complex time, $t=t_0+i\tau$.
The action $W_0$ in (\ref{A}) and (\ref{SFA-correct}) can be represented as
\beq
W_0(p_0)=\int\limits_{t_s}^{t_0}({\cal L}+E_0)dt-p_0b\, ,
\label{W0}
\eeq
where ${\cal L}=v^2/2-U(x)$ is the field-free Lagrange function.
Since the particle moves in complex time, $t\in[t_s,t_0]$, its velocity is imaginary, whereas the coordinate is real.
At the exit $x(t_0)=b$ all quantities become real.
To solve (\ref{NE0}), one has to consider $t_0$ as an external parameter and $\tau_0$ can be found from the initial condition $x_0(t_s=t_0+i\tau_0)=0$ (for an example see the next section).

After the exit, when time becomes real $t\ge t_0$, the particle moves under the action of the laser field.
The respective trajectory satisfies another Newton equation
\beq
\ddot x=\dot v=\dot p_F(t)\, ,~~~~v(t_0)=p_0\, ,~~~~x(t_0)=b
\label{NE1}
\eeq
with the solution
\beq
x(t)=b+p(t-t_0)+G(t)-G(t_0)\, ,~~~~~G(t)=\int\limits_0^tp_F(t^{\prime})dt^{\prime}\, .
\label{NE1-sol}
\eeq
The condition $\dot x(t_0)=p_0$ then specifies  $t_0$ to be a function of the final momentum, $p$.
The following algebra
\begin{eqnarray}
-S_0&=&-\int\limits_{t_0}^{+\infty}\left(v^2/2-E_0+\frac{d}{dt}vx-
\frac{d}{dt}vx\right)dt=
\nonumber \\
&=&\int\limits_{t_0}^{+\infty}\left(v^2/2+\dot vx+E_0\right)dt-vx\vert_{t\to +\infty}+vx\vert_{t=t_0}\, ,
\nonumber
\end{eqnarray}
allows to represent the action $S_0$ in a form identical to that of  (\ref{W0}).
Thus, the exponential in (\ref{MSFA-QS}) can be rewritten as $\exp (iW)$ where
\beq
W=\int\limits_{t_s}^{+\infty}({\cal L}+E_0)dt-vx\vert_{t\to+\infty}
\label{W}
\eeq
is the reduced action calculated along the classical complex trajectory selected described above. This is the basic result of the ITM \cite{popov-itm}.

The ITM provides a way to  generalize the amplitude (\ref{SFA-correct}). Indeed, one can calculate the function (\ref{W}) accounting for both the binding potential and the field of the electromagnetic wave, i.e. evaluating the trajectory $x(t)$ from the equation
\beq
\ddot x=\dot v=-{\partial U}/{\partial x}+\dot p_F(t)
\label{NE2}
\eeq
with initial and boundary conditions
\beq
x(t_s)=0\, ,~~~~~v^2(t_s)/2=E_0\, ,~~~~~v(t\to\infty)=p\, .
\label{ic}
\eeq
Except for the simplest model systems, a solution to Eqs.~(\ref{NE2}-\ref{ic}) can only be found numerically or by iteration with respect to one of the two fields. However, even in first-order perturbation theory, it is possible to account for the nonlinear effect of an intense laser field on tunneling.
Indeed, if in some part of the position space the laser field is small compared with the binding force (or vice versa), corrections to trajectory $\delta x$ and to the action $\delta W$ can be derived perturbatively.
These corrections must remain small compared to the respective unperturbed values, $|\delta W|\ll|W|$, to justify the application of perturbation theory. However, since under semiclassical conditions (\ref{K0}) the action is numerically large, the condition
\beq
 1 \ll|\delta W|\ll |W|
\label{deltaW}
\eeq
is typically fulfilled. This means that even a perturbative correction due to the presence of another (e.g. laser) field can cause a substantial modification of the spectra. The regime where such a semiclassical perturbation theory for the action is relevant results in a variety of strong field problems \cite{popov67,poprz-prl08,bauer08}.

The correction $\delta W$ consists of two parts: one due to the functionally different action that accounts for the additional interactions, and the other related to the modification of the trajectory. In the literature the first correction has been better studied than the second. In particular, in the work of Ivlev and Melnikov \cite{ivlev85} presenting the first ITM treatment of laser-assisted tunneling, only the first correction was accounted for.

To summarize, by taking into account both potential and laser fields, the differential probability of observing the electron with the asymptotic momentum $p$ takes the form (\ref{SFA-correct}) with the amplitude
\beq
M(p)=\sum_{\alpha}\frac{{\cal P}_0(v(t_{0\alpha}))\exp(iW(p,t_{s\alpha}))}{\sqrt{dp/dt_{0\alpha}+i\beta}}\, ,
\label{MITM-final}
\eeq
with $W(p,t_{s\alpha})$ calculated along Eqs.~(\ref{W})-(\ref{ic}). This distribution is the main result of the present work. It includes both the field-free and the laser-assisted tunneling and accounts for the redistribution of the particle momenta due to the laser field  after exiting the barrier.
It is relevant under conditions (\ref{K0}) with the additional requirement that the number of ATI-like peaks in the spectrum should be large to keep Eqs.~(\ref{dw-new}) and (\ref{regular}) valid, i.e. $p_0p_F\gg\omega$ or $z_F\gg 1$. Integration over the final momenta of the particle at the detector provides the total tunneling probability. In Appendix A it is shown that the field free decay rate follows from (\ref{MITM-final}) in the weak field limit $\mathcal{E}_0\to 0$.

A remaining important question is what happens to the spectrum for increasingly thick barriers up to the limit of a true bound state. For an infinitely thick barrier, the field-free decay vanishes and only ATI is possible. An examination of the equations of motion and the action considered above shows that those trajectories that correspond to the final energy $p^2/2>U_0$ survive for the infinite barrier and the respective ionization probability is nonzero. This will result in a common ATI spectrum (see also next Section). We can therefore state that the present formulation contains contributions from both  the laser-assisted and laser-induced processes. These contributions can be distinguished  according to the type of trajectories: trajectories that vanish for an infinitely thick barrier are responsible for LAT. One should note, however, that this classification is only qualitative, since both families of trajectories depend continuously on the barrier width.

%%%%%%%%%%%%%%%%%%%%%%%%%%%%%%%%%%%%%%%%%%%%%%%%%%%%%%%%%%%%%%%%%%%%%%%%%%%
\section{Rectangular barrier \label{sqbarrier}}
%%%%%%%%%%%%%%%%%%%%%%%%%%%%%%%%%%%%%%%%%%%%%%%%%%%%%%%%%%%%%%%%%%%%%%%%%%%

Here we consider an example of laser-assisted tunneling which admits an analytical solution. Let us assume a QS state in a 1D rectangular well formed by short range forces, as depicted in Figure~\ref{fig1}.
The central part of the well (where the particle remains trapped most of the time) is assumed to be narrow enough so that the external laser field cannot perform considerable work on the particle on this spatial scale, $\cale_0a\ll E_0$.
For simplicity we consider the limit $\kappa_0a\ll 1$ such that the width of the well\footnote{This should not be confused to the width of the barrier $b-a\approx b$, which is semiclassically large, $\kappa_0b\gg 1$.} does not enter the result. 

According to the ITM recipe, one has to find trajectories that start at $t=t_s$ inside the well $x(t_s)=0$, with the initial  kinetic energy  $v^2(t_s)/2=E_0-U_0\equiv -\kappa_0^2/2$ (so that $v(t_s)=i\kappa_0$) and arrive at the detector with the particle having the desirable final momentum $p$. The detailed derivation is presented in Appendix B. For the monochromatic field (\ref{et}) all periods are equivalent, so that there are only two essentially different solutions, all the others being obtained by a $2\pi j$ translation. The summation over all periods gives the factor $\omega\, \delta(p^2/2-E_0+U_P-j\omega)$, which expresses energy conservation. The initial imaginary time $\omega t_s=\varphi_s=\phi_0+i\psi_0$ satisfies the equations (see also Appendix B):
\begin{eqnarray}
b&=&(\kappa_0/\omega)\psi_0-(p_F/\omega)\cos\phi_0(\cosh\psi_0-1-\psi_0\sinh\psi_0)\, ,
\nonumber \\
p&=&v_0(\varphi_s)+p_F\sin\phi_0\, .
\label{speq-3}
\end{eqnarray}

The solution of these equations and the respective trajectories given by (B.3) and (B.4) enter the expression of the action $W(p,t_{s\alpha})$.
The differential decay rate $R$ over the whole observation time $T$ is  given by
\begin{equation}
dR=\frac{dw(p)}{T}=\sum_{j}\omega^2\frac{\delta(p-p_j)}{2\pi p_j}\frac{8\kappa_0^3p_0}{\kappa_0^2+p_0^2}
\left\vert\sum_{\alpha=1,2}\frac{\exp\left(iW(p,t_{s\alpha})\right)}
{\sqrt{dp/dt_{0\alpha}+i\beta}}\right\vert^2dp\, ,
\label{therate}
\end{equation}
where $p_j=\sqrt{p_0^2-p_F^2/2+2j\omega}$ are the momenta corresponding to the ATI-like peaks.
Here we have taken into account that the prefactor $\mathcal{P}_0$ corresponding to the field-free rate $R_0=\mathcal{P}_0^2 \mathrm{exp}(-2\kappa_0 b)$ is given (in the narrow well limit $\kappa_0a\ll 1$) by
\begin{equation}
\mathcal{P}_0^2=\frac{8\kappa_0^3p_0}{\kappa_0^2+p_0^2}\, .
\end{equation}

In the weak field regime the result for the action can be simplified up to a short analytic form by keeping only linear  terms in $\cale_0$.
For a trajectory $\alpha$, this procedure gives
\begin{eqnarray}
W_{\alpha}(p)&=&i\kappa_0 b\bigg(1+\frac{\cale_0b\cos\phi_{0\alpha}}{2\kappa_0^2}+\frac{p_0\cale_0\psi_{00}^3\sin\phi_{0\alpha}}{3\kappa_0 b\omega^2}\bigg)+\frac{p_0(p-p_0)}{\omega}\phi_{0\alpha} \nonumber \\
&+&\psi_{00}\frac{\cale_0b^2\sin\phi_{0\alpha} }{2\kappa_0}-pb
+p_Fb\sin\phi_{0\alpha}+\frac{\kappa p_F}{\omega}\cos\phi_{0\alpha}\, ,
\label{W-PT}
\end{eqnarray}
with $\psi_0\approx\psi_{00}=b\omega/\kappa_0\ll 1$.
From this expression one derives the true criterion for the weak-field limit corresponding to the ``exclusive'' regime,
\beq
\mu=\kappa_0 b\frac{\cale_0b}{\kappa_0^2}\ll 1~.
\label{mu-new}
\eeq
Under the further simplification that only the most probable trajectory is considered (i.e. $\phi_0=0$ or $\pi$) and the real part of the action is disregarded (no interference), the result of Ivlev and Melnikov is recovered \cite{ivlev85}.
In this case, however, the result of \cite{ivlev85} is more general than the one derived in this section, since the former does not assume the potential to be a rectangular barrier.

For the action (\ref{W-PT}), the probability vanishes in the limit $b\to\infty$. This, however, cannot be consistent with the existence of ATI for an infinitely thick barrier. This apparent paradox can be solved if we take into account that the results of this subsection (including also Appendix B) are derived  assuming that the particle escapes from under the barrier at its right edge, $x=b$. This is, however, true only when $\cale_0b\le U_0-E_0$ (see Appendix B). Clearly, with increasing barrier thickness $b$, this condition  will be violated for any given field amplitude and trajectories that escape through the tilted part of the barrier will come into play. Along such trajectories, the action  becomes $b$-independent and virtually identical to the case of common ATI.

%%%%%%%%%%%%%%%%%%%%%%%%%%%%%%%%%%%%%%%%%%%%%%%%%%%%%%%%%%%%%%%%%%%5
\subsection{Numerical results}
%%%%%%%%%%%%%%%%%%%%%%%%%%%%%%%%%%%%%%%%%%%%%%%%%%%%%%%%%%%%%%%%%%%

As numerical examples we first consider tunneling through two rectangular barriers of parameters $U_0=3.0,~b=3.0$ and $U_0=4.0,~b=10.0$ assisted by a laser field with frequency $\omega=0.1$. The action $W$ and the rate $R$ are obtained using the expressions given in Appendix B and Eq.~(\ref{therate}). In Figure~\ref{fig2} we present the imaginary part of the action  and the spectrum of the laser-assisted tunneling as a function of the final energy $E$ for three different field amplitudes $\cale_0$. For each energy between $E_{\rm min}$ and $E_{\rm max}$ in Figure~\ref{fig2}(a) and (d) the imaginary part of the action $\rm{Im}(W)$ has two values corresponding to the two trajectories starting  inside the well at times $\varphi_{s\alpha}$ with $\alpha=1,2$.  The spectrum consists of several ATI-like maxima whose positions are dictated by the energy conservation conditions in Eq.~(\ref{therate}).

\begin{figure}
\includegraphics[width=8cm]{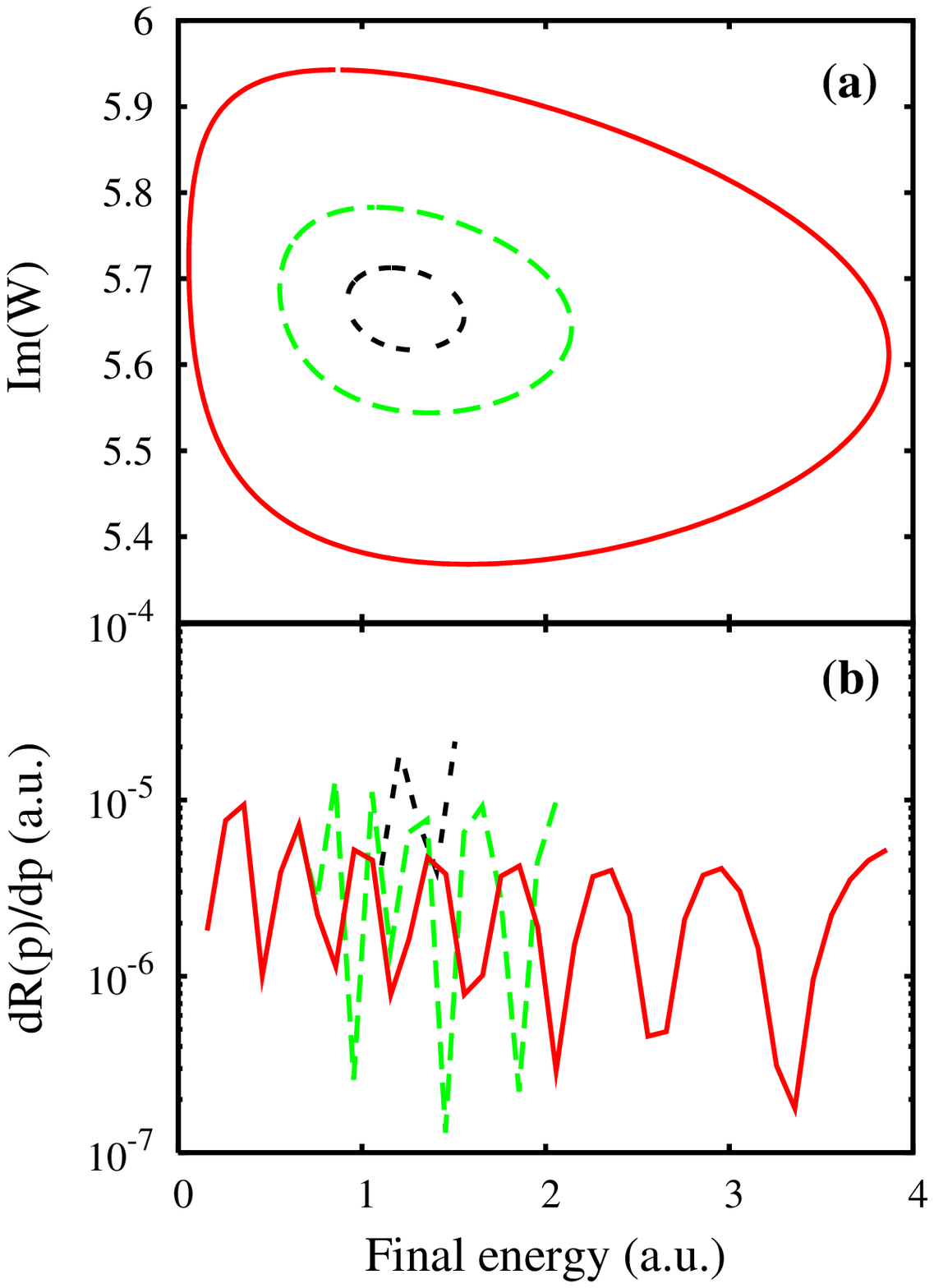}
\includegraphics[width=8cm]{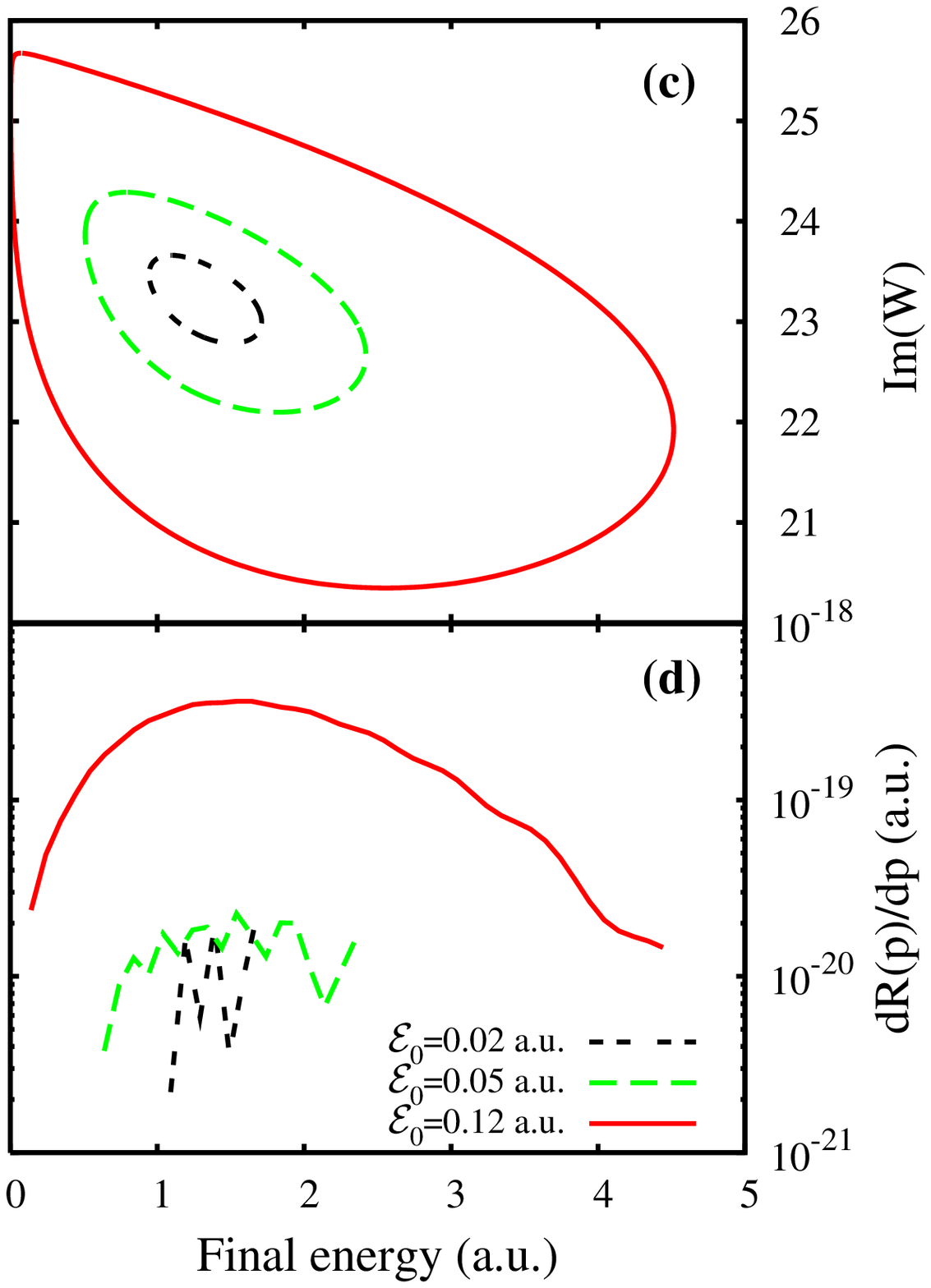}
\caption{Imaginary part of the action ${\rm Im}(W)$ (a), (c) and  the spectra (b), (d) as a function of the the final energy $E=p^2/2$ for two sets of parameters: $U_0=3.0$, $b=3.0$, $E_0=1.217$, $\omega=0.10$, (a)-(b) and $U_0=4.0$, $b=10.0$, $E_0=1.302$, $\omega=0.10$, (c)-(d). Each panel shows three curves for $\cale_0=0.02$ (black double-dashed line), $\cale_0=0.05$ (dashed green line) and $\cale_0=0.12$ (solid red line). }
\label{fig2}
\end{figure}

In the case of the thin barrier with $b=3$, the tunneling occurs mostly field-free, and the main effect of the laser is to change the particle momentum  after the barrier exit. By comparing the results for different field intensities in Figure~\ref{fig2}b, we observe that the spectra become more narrow with decrease of $\mathcal{E}_0$, approaching the  Lorenzian shape of the field-free decay rate. Furthermore,  the ratios between the field-free and laser-induced tunneling probabilities  do not vary much for the considered field amplitudes $R_{0.02}/R_0=0.41$, $R_{0.05}/R_0=0.62$ and  $R_{0.12}/R_0=0.89$.  Note that for these parameters, the large number of ATI-like maxima condition $p_0p_F\gg\omega$, necessary to justify the momentum distribution along Eq.~(\ref{dw-new}) is not fulfilled.  As a consequence, our distribution loses a part of the tunneled particles, and the laser-assisted to field-free decay ratios are less than unity. Due to this effect, it is more informative to consider as reference the laser-assisted rate at the lowest field intensity. This gives us $R_{0.12}/R_{0.02}\approx 2$, showing that the laser field of the amplitude $\cale_0=0.12$ only enhances the total decay rate by the factor of 2 for this barrier width.

In contrast, for the thick barrier of width $b=10$, corresponding to a very small field-free decay rate, the ``inclusive'' regime is achieved, in which the laser  has a substantial effect also on the tunneling rate itself. Here, the ratios of the field-free and laser-induced tunneling probabilities are $R_{0.02}/R_0=0.72$, $R_{0.05}/R_0=1.94$ and  $R_{0.12}/R_0=58.17$, extending over almost two orders of magnitude with the increasing field.  Thus, as anticipated in the introduction,  our approach reproduces important qualitative conclusions of earlier studies, namely that depending on the parameters, there are two different regimes of decay: ``exclusive'',  when the spectrum is strongly affected without a modification of the total decay rate (the thin barrier) and ``inclusive'',  when the rate of decay is strongly modified by the laser field (the thick barrier). Quantitatively, these two regimes can be distinguished according to value of the parameter (\ref{mu-new}). The ``exclusive'' regime requires $\mu\ll 1$ and holds for the first set of parameters, while for the thick barrier we enter the ``inclusive'' domain, $\mu\gg 1$.

\begin{figure}
\includegraphics[width=8cm]{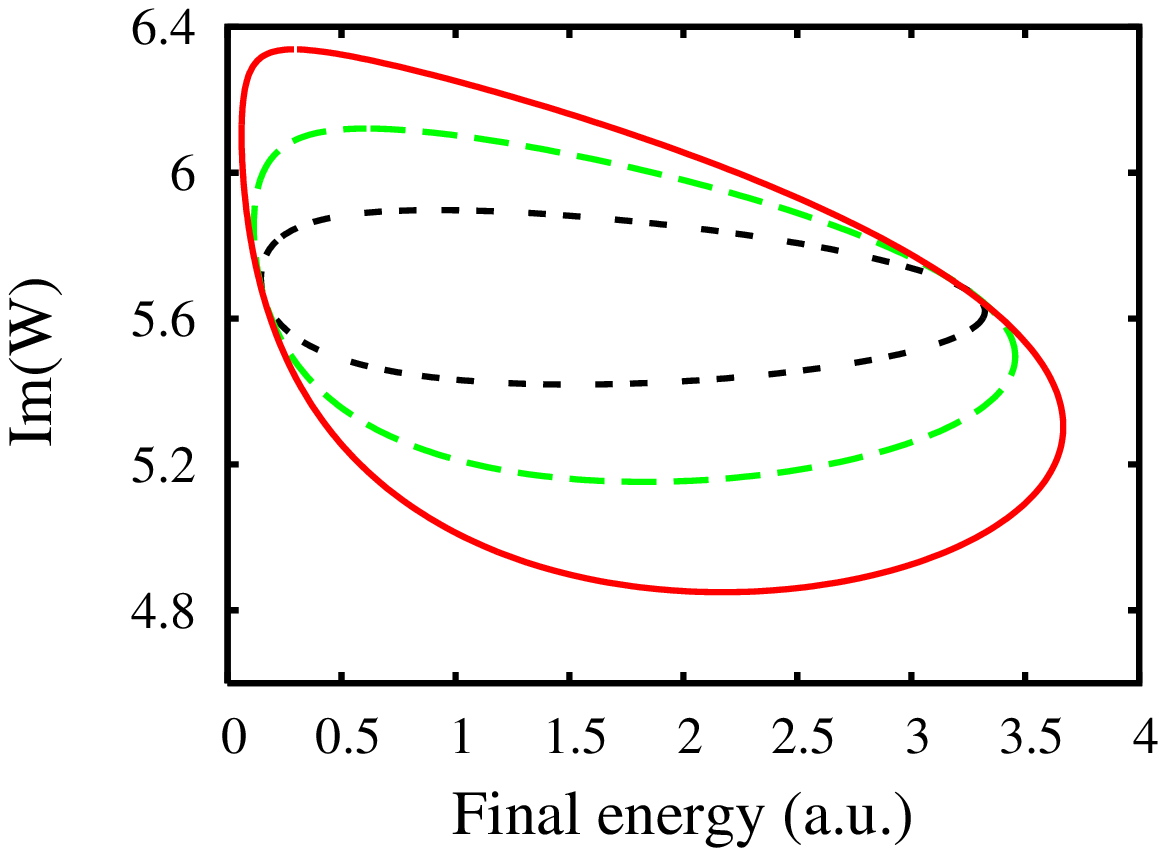}
\includegraphics[width=8cm]{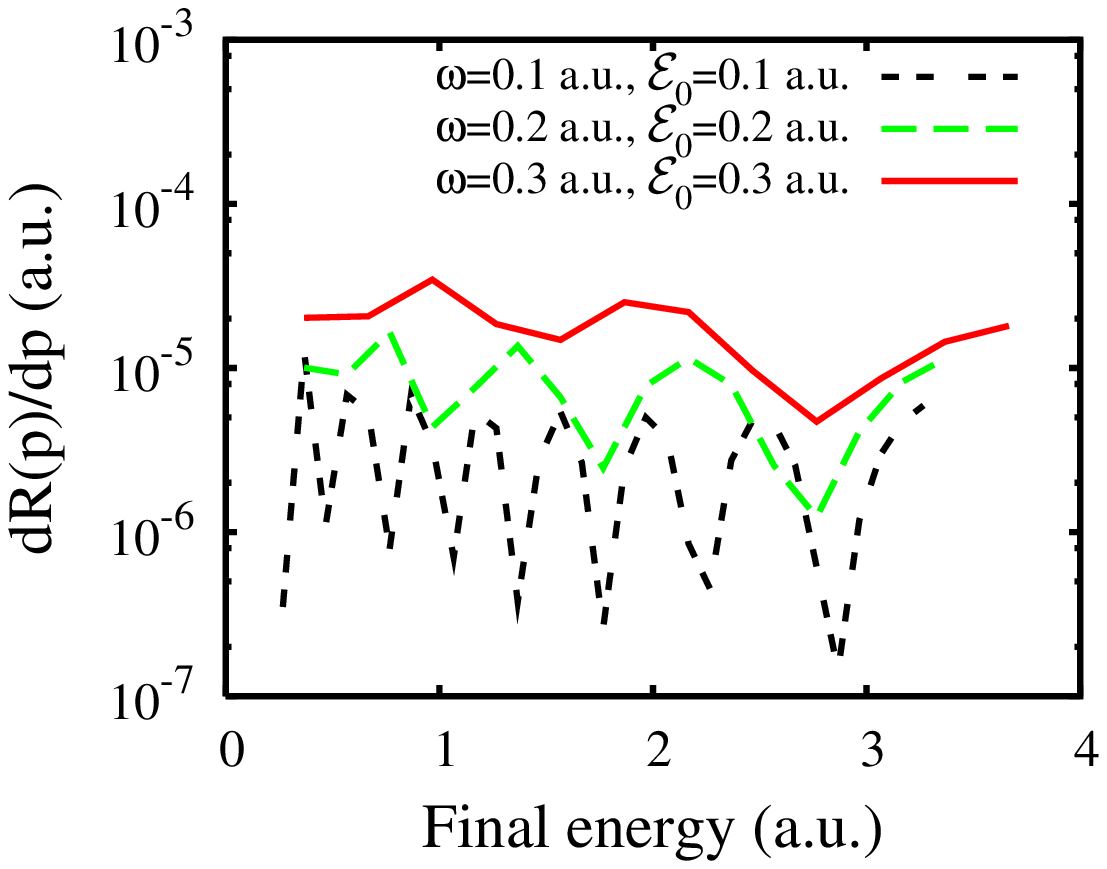}
\caption{Imaginary part of the action (left) and laser assisted decay spectra (right) calculated for  monochromatic fields with various frequencies and  field momenta $p_F=1$    for the thin barrier considered in Figure~\ref{fig2}. The field amplitudes and frequencies are $\cale_0=0.1$  and $\omega=0.1$  (black double-dashed line), $\cale_0=0.2$  and $\omega=0.2$  (green dashed line) and $\cale_0=0.3$  and $\omega=0.3$  (red solid line), respectively.}
\label{fig3}
\end{figure}

In Figure~\ref{fig3} we have considered the case of the thin barrier with $b=3$ for different laser field amplitudes and frequencies such that the field momenta are $p_F=1$. The ratio of the different field frequencies can be identified from the spectra. The total tunneling rate is increasing with increasing the field, with  $R_{0.1}/R_0=0.74$, $R_{0.2}/R_0=0.91$ and  $R_{0.3}/R_0=1.56$.

\begin{figure}
\includegraphics[width=8cm]{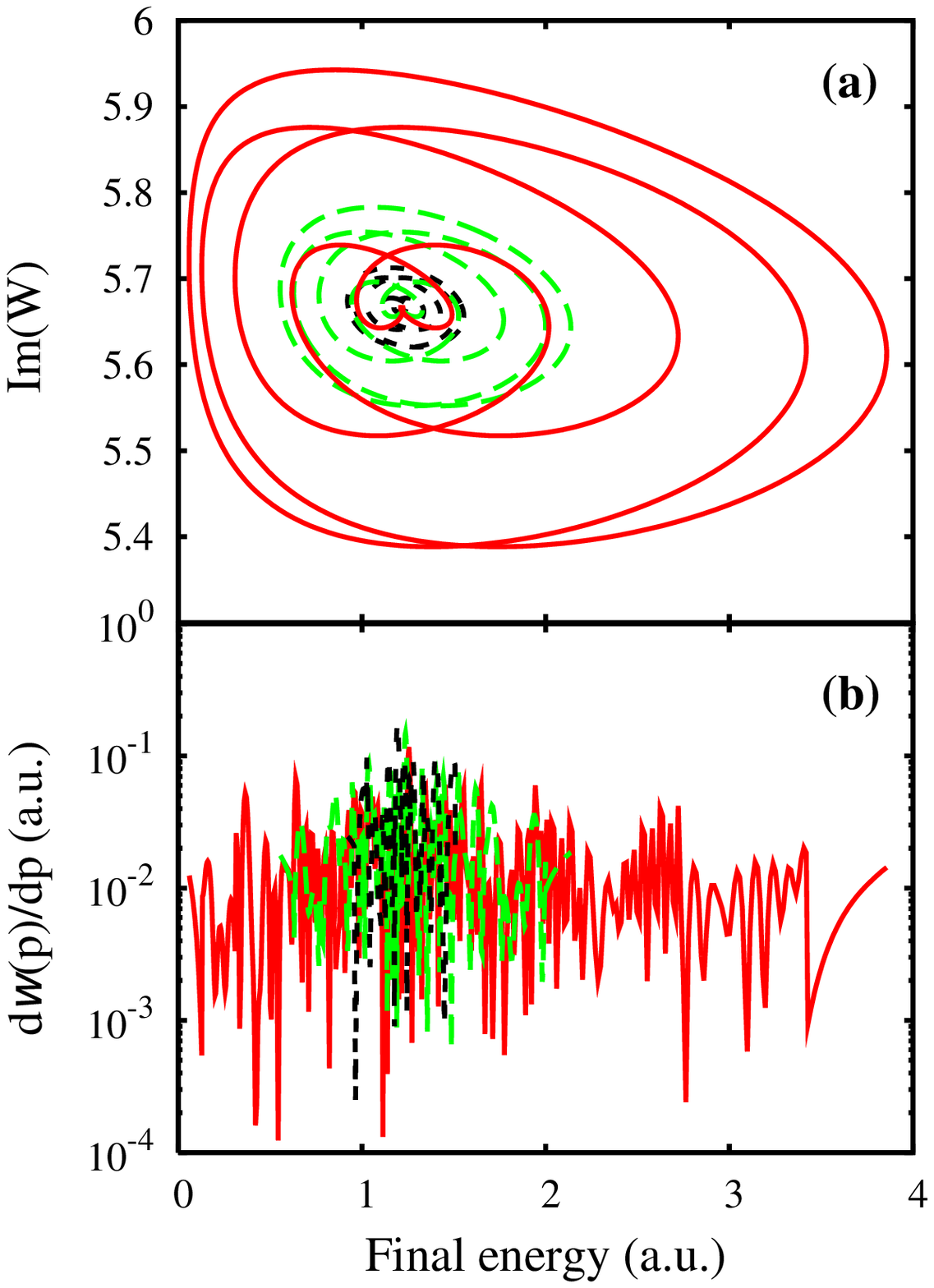}
\includegraphics[width=8cm]{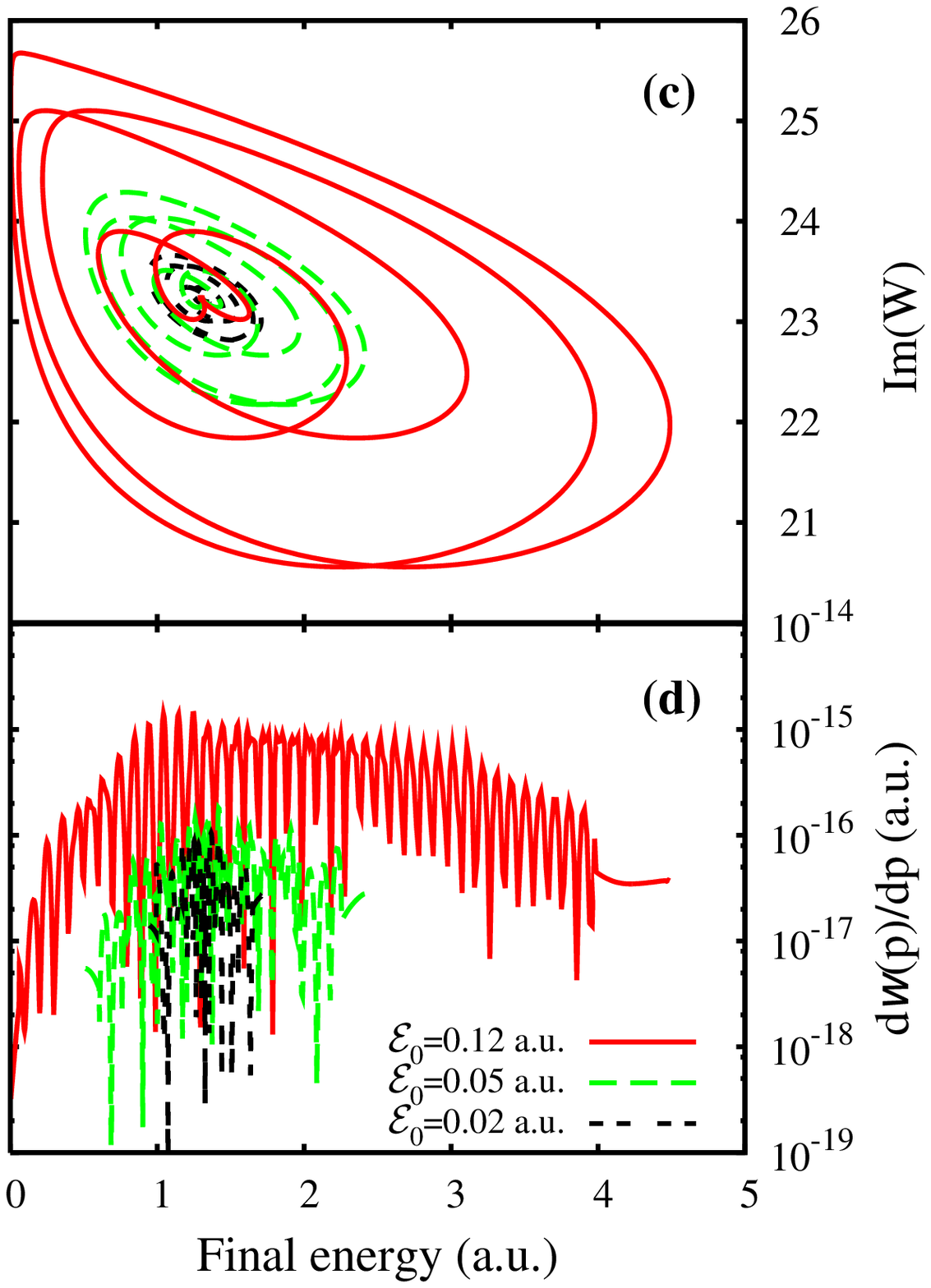}
\caption{Imaginary part of the action (a), (c) and laser assisted decay spectra (b), (d) calculated for a finite pulse of the form (\ref{shortpulse}) for the two barrier parameter sets of Figure~\ref{fig2} and fields of frequency $\omega=0.1$  and  amplitudes $\cale_0=0.02,~0.05,~0.12$. }
\label{fig4}
\end{figure}

If we consider instead of the monochromatic field a few-cycle laser pulse, the spectrum of Dirac-delta functions will be replaced by a comb of broadened ATI-like maxima between $E_{\rm min}$ and $E_{\rm max}$. Spectra obtained using the extended ITM for a six-cycle pulse of the form 
\beq
\cale(t)=\cale_0\sin^2\left(\frac{\omega t}{2n_p}\right)\cos\omega t\, ,~~~~n_p=6
\label{shortpulse}
\eeq
for the parameters previously addressed in the text are presented in Figure~\ref{fig4}. 
Here the amplitude consists of up to $2n_p$ coherent contributions that produce the interference pattern of the spectra.
The broad ATI-like maxima can be observed best for the thick barrier case in Figure~\ref{fig4}d, where the absolute values of the two contributions from a given laser period differ substantially and smear out the interference. Unlike the case of tunneling assisted by a monochromatic field where the differential rate $dR(p)$ is calculated, the spectra in Figure~\ref{fig4} present the differential probability $dw(p)/dp$. The ratio $w(p)/$pulse duration delivers approximately the decay rate.

%

%%%%%%%%%%%%%%%%%%%%%%%%%%%%%%%%%%%%%%%%%%%%%%%%%%%%
\subsection{Comparison with numerical results of the time-dependent Schr\"odinger equation  (TDSE)\label{comparison}}
%%%%%%%%%%%%%%%%%%%%%%%%%%%%%%%%%%%%%%%%%%%%%%%%%%
 
In order to check the accuracy of our approach we compared the obtained spectra with accurate numerical results of the TDSE. For the test case of the rectangular barrier, we have calculated  exact numerical spectra using a 1D version of the QPROP Schr\"odinger solver \cite{QPROP}, which propagates the wavefunction in real time on a spatial grid. For the numerical simulations, the width of the potential well (in which $U=0$) needed to be specified and was chosen as $a=\pi/2$. We first obtained the ground state wavefunction for the particle inside the well of height $U_0$=3 and $b\to\infty$, i.e. a barrier of infinite width, as depicted by the dashed line in Figure \ref{fig1}. The ground state energy on the numerical grid of spacing $\Delta x=0.1$ was  $E_0=1.24$. At time $t=0$, the infinite barrier was replaced by a barrier of finite width $b-a=4$, keeping the height constant. As a consequence, tunneling occurs for $t>0$. The sudden switch from the infinite to the finite barrier disturbs the trapped electron and leads to a short time interval of transient tunneling dynamics before a constant free-tunneling rate is established. However, this time interval of a few atomic units is much shorter than the pulse duration so that it did not make a difference whether the six-cycle laser pulse of the form (\ref{shortpulse}) and frequency $\omega=0.057$  (Ti:Sa laser) was switched on at $t=0$ or with a delay. In any case, the numerical grid was big enough to support the entire wavefunction during the propagation time without reflections off the grid boundary. The electron spectra were calculated using the window operator technique (see, e.g., \cite{QPROP}) and normalized to the field-free decay.

\begin{figure}
\includegraphics[width=8cm]{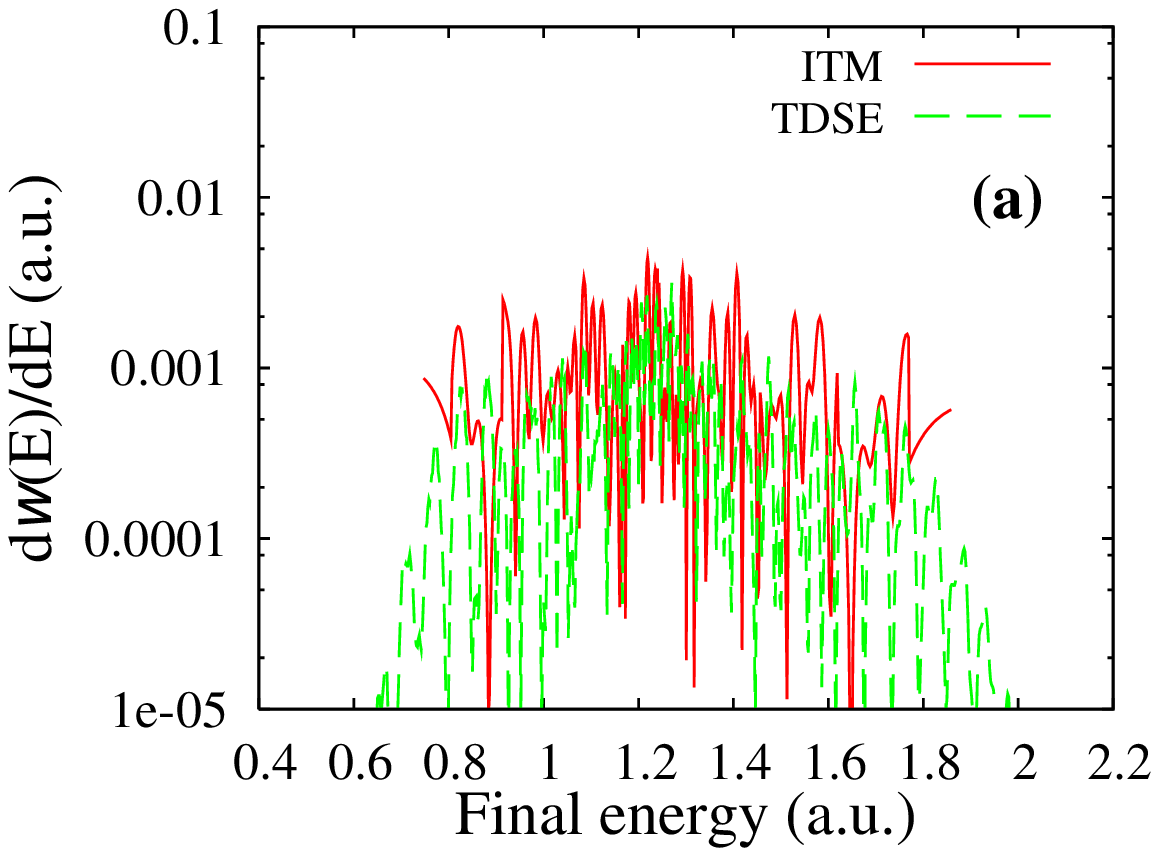}
\includegraphics[width=8cm]{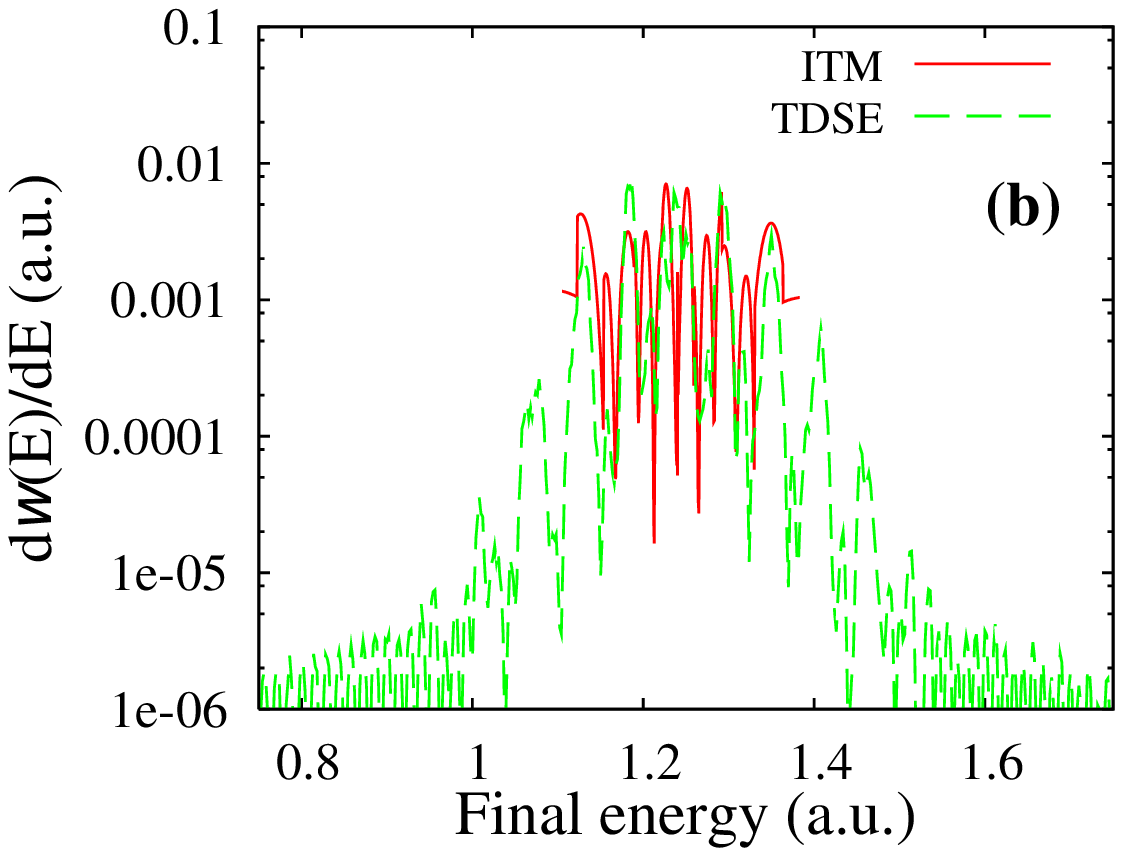}
\includegraphics[width=8cm]{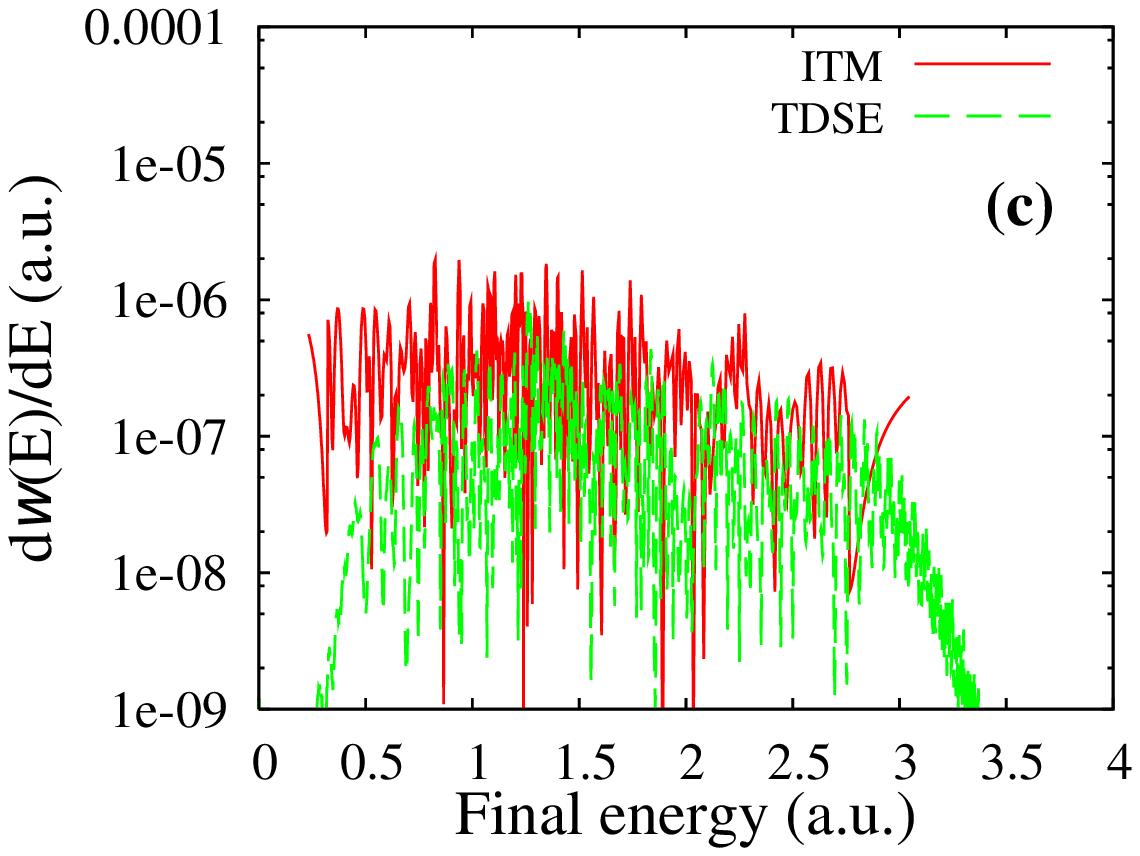}
\includegraphics[width=8cm]{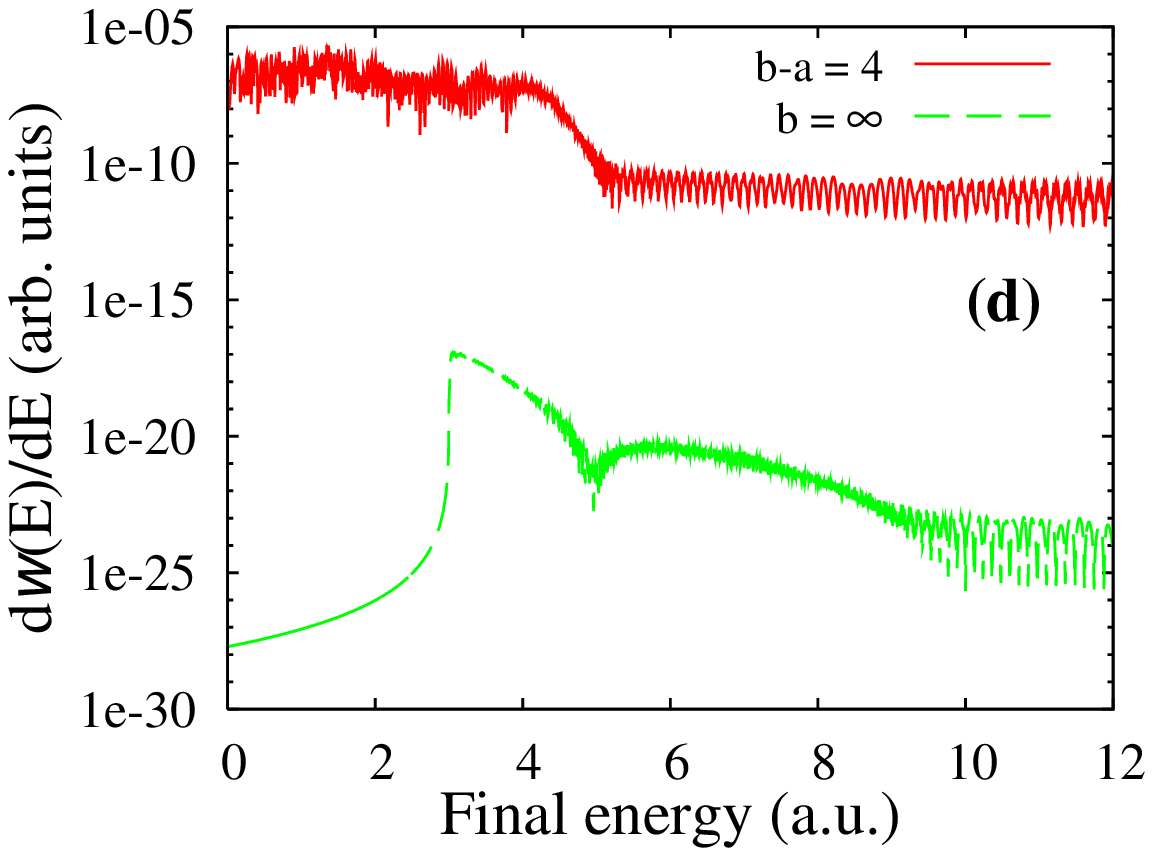}
\caption{(a),(b),(c) Tunneling probabilities $dw(E)/dE$ for rectangular barriers of widths $b-a=4$  (a), (b), and  $b-a=6$ (c) as a function of the final energy: the present ITM calculation (solid red line) and numerical solution of the TDSE (green dashed line). The laser field parameters are  $\omega=0.057$  and  (a) $\mathcal{E}_0$=0.02, (b) $\mathcal{E}_0$=0.005  and (c) $\mathcal{E}_0$=0.05. (d) Comparison between laser-assisted tunneling through the finite barrier considered in (a) under the action of a field of amplitude $\mathcal{E}_0$=0.075 (solid red line) and ATI for $b=\infty$ in the same field (green dashed line).
 }
\label{fig5}
\end{figure}

In Figure~\ref{fig5}, we compare the obtained tunneling probabilities for a  barrier of thickness $b-a=4$, height $U_0=3$  and initial particle energy $E_0=1.24$  under the action of fields of amplitudes $\mathcal{E}_0$=0.02  (Figure~\ref{fig5}a) and $\mathcal{E}_0$=0.005   (Figure~\ref{fig5}b). The spectra agree well both qualitatively and quantitatively for final energies within the CB, with the ITM results slightly higher than the TDSE ones. From a comparison of the two field-free decay rates, we observe that the fitted from the TDSE results are always smaller than the calculated $R_0$. This behavior is related to TDSE numerics requirements, which cannot handle the very thin potential well limit $a\to 0$.

The ITM approach delivers spectra that  vanish abruptly beyond the CB and cannot reproduce the shoulders visible in the TDSE results. As already discussed in Section~\ref{SFA-various}, the saddle-point method  is actually not applicable in the form described here outside the CB. The correct approach requires us to include the term proportional to the third derivative of the action in the phase decomposition near the saddle point \cite{gor99}. For broad spectra, the contribution of the shoulders outside the CB is not significant and our approach provides reliable results. The contribution of the shoulders increases for narrow spectra, as one can see comparing Figures \ref{fig5}a and \ref{fig5}b. 

We further compare the ITM and TDSE results for a thicker barrier with $b-a=6$  and field amplitude $\mathcal{E}_0$=0.05  in Figure~\ref{fig5}c. Here the agreement is less accurate at small energies $E<0.8$, where the ITM results are about one order of magnitude higher than the TDSE ones. Since ITM delivers a momentum spectrum, the variable transformation $dE=pdp$  introduces a divergence for asymptotic momenta approaching the origin. However, for energies close to the initial particle energy and larger, the ITM agrees well with the TDSE. We conclude that the ITM provides not only qualitative but also quantitative results for laser-assisted tunneling of QS states within the semiclassical parameter regime.

Note that in all numerical examples considered in this section, two effects were disregarded in the ITM. Firstly, in our analytical formulae  trajectories that escape through the tilted barrier are not accounted for, so that the contribution that turns in the limit $b\to\infty$  into ATI is missing. For the parameters we have chosen, this contribution can be safely neglected. For demonstration, we show in Fig.~\ref{fig5}d the LAT spectrum calculated numerically using the TDSE for the barrier parameters of panel (a) under the action of a laser field of amplitude $\mathcal{E}_0=0.075$ and compare it with ATI through an infinite barrier ($b=\infty$) of the same height. We can see that for the chosen parameters,  the ATI probability  is many  orders of magnitude smaller than laser-assisted tunneling one. Secondly, the theory does not take into account that the electron can be driven back to the barrier and rescatter absorbing or emitting additional photons. As is known from the literature, rescattering leads to the formation of one or more plateaus in the spectrum, with the characteristic number of peaks given by the Reiss parameter (\ref{zF}).  The same effect can also be interpreted as multiphoton stimulated bremsstrahlung \cite{bunkin66}. For our calculations, we have selected the parameters such that in all cases $p_0>p_F$ and rescattering plays no significant role. The rescattering plateau can be reproduced by TDSE calculations covering the higher field amplitude domain $p_F>p_0$, which was, however, not considered in this work.

\section{Conclusions}
%%%%%%%%%%%%%%%%%%%%%%%%%%%%%%%%%%%%%%%%%%%%%%%
 Starting from the SFA and its formulation in terms of complex trajectories, we have developed a general method to describe the laser-assisted decay of QS states in the semiclassical parameter regime. In order to illuminate the physical essence of the problem and avoid unnecessary technical complications,  we have neglected some accompanying effects like conventional ATI and rescattering. A test case comparison with numerical results of the Schr\"odinger equation shows not only qualitative but also quantitative agreement. 

Due to the general statement of the problem, our method can be applied with moderate numerical effort to  many laser-assisted tunneling processes within the semiclassical regime. 
In particular, laser-assisted tunneling through a more realistic Coulomb barrier as it occurs in autoionization, autoemission  or generation of high harmonics in the presence of static or low-frequency fields \cite{starace} can also be described via our extended ITM. Here the real strength of the ITM could be used to correctly take into account the effect of long-range potentials that are acting on the  particle in the classically allowed domain simultaneously with the laser field. Work in this direction is in progress.

As a further  incentive, recently ATI was experimentally observed at sharp metal tips driven by moderately intense infrared laser pulses in the presence of a DC field \cite{schenk}.
The latter formed an effective tilted barrier that the electrons tunneled through with the assistance of the laser field.  Under the action of the AC laser field, the electron current was modulated in time, forming femto- and attosecond dense electron bunches \cite{papadichev}. An extension beyond the 1D case of the method  developed in this work can be applied to model such experiments. For non-separable variables in many dimensions, our approach is expected,  however, to involve more cumbersome calculations (see also Ref.~\cite{KapurPeierls}), since the corresponding multidimensional trajectories and complex times can only be found numerically. This should, however, remain an essentially simpler task as compared to obtaining the numerical solution of the corresponding TDSE.

\section*{Acknowledgements}
%%%%%%%%%%%%%%%%%%%%%%%%%%%%%%
We would like to thank  C. H. Keitel, W. Becker, V. D. Mur and V. S. Popov for fruitful discussions. AP acknowledges financial support of the Robert-Bosch-Stiftung  in the framework of the ``Fast Track" Programme.
SVP acknowledges financial support from the Deutsche Forschungsgemeinschaft (project No. SM 292/1-1) and from the Russian Foundation for Basic Research (project No. 11-02-91331). DB acknowledges support from the Deutsche Forschungsgemeinschaft (SFB 652).

%%%%%%%%%%%%%%%%%%%%%%%%%%%%%%%%%%%%%%%%%%%%%%%%%%%%%%%%%%%%%%%%%
\appendix
\section{Field-free limit}
%%%%%%%%%%%%%%%%%%%%%%%%%%%%%%%%%%%%%%%%%%%%%%%%%%%%%%%%%%%%%%%%%
In this Appendix we show that in the weak field limit the field-free decay rate
\begin{equation}
R_0={\cal P}_0^2\exp(-2{\rm Im}W_0)
\label{fieldfreerate}
\end{equation}
follows from the amplitude (\ref{MITM-final}). Quantitatively, the weak field limit is determined by the condition (\ref{mu-new}). Correspondingly,  the subbarrier correction to the action is smaller than unity so that the imaginary part of the action is given by the field-free contribution $W_0$ (\ref{W0}).
Since the momentum change during the subbarrier motion is also small,  the initial velocity at the exit is $v(t_0)=p_0$ and then the final momentum is given by
\begin{equation}
p(t_0)=p_0+p_F\sin\omega t\, .
\end{equation}
We have assumed here a monochromatic field (\ref{et}) for simplicity. The spectrum consists of $L=2p_0p_F/\omega\gg 1$ ATI-like peaks with energies between $(p_0+p_F)^2/2$ and $(p_0-p_F)^2/2$. For the given momentum inside this interval, there are two solutions per period, so that
\begin{eqnarray}
\omega t_{0n}^{(+)}&=&{\rm arcsin}\bigg(\frac{p-p_0}{p_F}\bigg)+2\pi n\, , \nonumber \\
\omega t_{0n}^{(-)}&=&\pi-{\rm arcsin}\bigg(\frac{p-p_0}{p_F}\bigg)+2\pi n\, . 
\end{eqnarray}
In the limit we consider, the laser field only enters the action via these initial times $t_0$,
\begin{eqnarray}
W&=&W_0+\int\limits_{t_0}^T(p^2/2+E_0)dt-px(T)+p_0b \nonumber \\
&=&i{\rm Im}W_0+\frac{1}{2}(p_0^2-p^2)(T-t_0)\, ,
\end{eqnarray}
where $T$ is the large observation time and we take into account that $x(t)\approx b+p(t-t_0)$.
The sum over the laser periods gives
\begin{equation}
\sum_{n=0}^N\exp\bigg(i(p^2-p_0^2)\frac{\pi n}{\omega}\bigg)\to\sum_j\delta\bigg(\frac{p^2-p_0^2}{2\omega}-j\bigg)~,~~~~N\to\infty~.
\end{equation}
Then
\begin{eqnarray}
dR=\frac{dw}{T}&=&\frac{{\cal P}_0^2(p_0)}{2\pi p_F}\exp(-2{\rm Im}W_0) \nonumber \\
&\times &\sum_j\delta\bigg(\frac{p^2-p_0^2}{2\omega}-j\bigg)\bigg\vert\frac{\exp(ij\omega t_0^{(+)})}{\sqrt{\cos\omega t_0^{+}}}-\frac{\exp(-ij\omega t_0^{(+)})}{\sqrt{-\cos\omega t_0^{+}}}\bigg\vert^2dp\, .
\end{eqnarray}
Under the condition $p_0p_F/\omega\gg 1$, the number of ATI-like peaks is large and the sum over $j$ can be replaced by an integral which evaluates to 1. The resulting distribution should be integrated over $dp$ within the limits $p_0\pm p_F$. Taking into account that $\cos\omega t_0^{+}=\sqrt{1-(p-p_0)^2/p_F^2}$ and disregarding the rapidly oscillating interference term in the modulus square, we obtain precisely the field-free rate (\ref{fieldfreerate}).

%%%%%%%%%%%%%%%%%%%%%%%%%%%%%%%%%%%%%%%%%%%%%%%%%%%%%%%%%%%
\section{ Rectangular barrier}
%%%%%%%%%%%%%%%%%%%%%%%%%%%%%%%%%%%%%%%%%%%%%%%%%%%%%%%%%%%
Here we present in more detail the ITM treatment for the case of laser-assisted tunneling through a rectangular barrier discussed in Section~\ref{sqbarrier}. In the field-free case, to each initial time instant $t_0$ corresponds the trajectory
\begin{eqnarray}
x_0(\varphi)&=&i(\kappa_0/\omega)(\varphi-\varphi_s)\, ,~~~~\psi\in [\psi_0,0]\, ,~~~~\psi_0=b\omega/\kappa_0\, ,
\nonumber \\
x_0(\varphi)&=&b+(p_0/\omega)(\varphi-\phi_0)\, ,~~~~\varphi\in [\phi_0,+\infty)\, .
\end{eqnarray}
Here and below we use the dimensionless time $\varphi=\omega t=\phi+i\psi$ such that $\varphi_s=\phi_0+i\psi_0$. The field-free action is then
\begin{equation}
W_0=ib\kappa_0-p_0b\, .
\end{equation}
The trajectory is always real, while the velocity is imaginary in complex time and becomes real when time is real at $x\ge b$. This allows for the standard interpretation that the particle moves under the potential barrier having an imaginary velocity. Later on, the particle escapes from under the barrier and its trajectory becomes entirely classical. As it will be seen in the following, for the time-dependent case this subdivision no longer holds: a trajectory can have a nonzero imaginary part even in real time.
However, for the sake of convenience, we will still refer to subbarrier motion when time is complex, $\psi\ne 0$ and to  motion after the barrier otherwise.

In the presence of the field (\ref{et}), the trajectory under the barrier is given by
\begin{eqnarray}
v^{\rm I}(\varphi)&=&\kappa_0[i-\gamma^{-1}(\sin\varphi-\sin\varphi_s)]\, ,
\nonumber \\
x^{\rm I}(\varphi)&=&\frac{\cale_0}{\omega^2}[(i\gamma+\sin\varphi_s)(\varphi-\varphi_s)+(\cos\varphi-\cos\varphi_s)]\, .
\label{trajectory1}
\end{eqnarray}
After the barrier, $\varphi\ge\phi_0$, where $U(x)=0$
\begin{eqnarray}
v^{\rm III}(\varphi)&=&v_0-p_F(\sin\varphi-\sin\phi_0)~,~~\mathrm{with} ~p=v_0+p_F\sin\phi_0\, ,
\nonumber \\
x^{\rm III}(\varphi)&=&x^{\rm I}(\phi_0)+p/\omega(\varphi-\phi_0)+(p_F/\omega)(\cos\varphi-\cos\phi_0)\, .
\label{trajectory2}
\end{eqnarray}
Here we introduce the dimensionless parameter
\begin{equation}
\gamma=\kappa_0\omega/\cale_0\,
\end{equation}
having the same physical meaning as the Keldysh parameter \cite{keldysh}, i.e. the ratio of the ``atomic'' momentum $\kappa_0$ to the field-induced momentum $p_F=\cale_0/\omega$.

Trajectories (\ref{trajectory1}) and (\ref{trajectory2}) are, however, insufficient to find the initial complex time, because the velocity is discontinuous at the barrier exit, $x=b$.
To avoid this discontinuity we replace the step-like potential by a smooth one; for example, the one depicted in Figure ~\ref{fig1} by the dashed-dotted line. The potential then drops from $U_0$ linearly down to zero on the width $\Delta$. This triangular barrier we denote as region II, while outside the barrier we define region III. In the intermediate region II the solution is
\begin{eqnarray}
v^{\rm II}(\varphi)
&=&v^{\rm I}(\varphi^{\prime})-p_F(\sin\varphi-\sin\varphi^{\prime})+(F_0/\omega)(\varphi-\varphi^{\prime})\, ,~~~~F_0=U_0/\Delta\, ,
\nonumber \\
x^{\rm II}(\varphi)&=&x^{\rm I}(\varphi^{\prime})+(p_F/\omega)(\cos\varphi-\cos\varphi^{\prime})+(F_0/2\omega^2)(\varphi-\varphi^{\prime})^2\, .
\end{eqnarray}
Here $\varphi^{\prime}$ is the time instant when ${\rm Re}[x^{\rm I}(\varphi^{\prime})]=b$. Matching the solutions in the domains I, II and III, we can find the constant $v_0$ as a function of the other parameters. Note that the sharper the slope is, the closer the time instant $\varphi^{\prime}=\phi_0+i\psi^{\prime}$ is to $\phi_0$. Decomposing the equations with respect to $\psi^{\prime}\ll 1$ we obtain:
\begin{eqnarray}
v_0&=&\left\{ p_0^2+2\kappa_0p_F\cos\phi_0\sinh\psi_0 \right. \nonumber \\
&+& \left. p_F^2[(\cosh\psi_0-1)^2\sin^2\phi_0-\cos^2\phi_0\sinh^2\psi_0]\right\}^{1/2}\, .
\end{eqnarray}
The two equations determining the initial complex time $\varphi_s$ follow from the requirements $x(\phi_0)=b$ and $v(\phi\to\infty)=p$ and take the form (\ref{speq-3}). They can be solved analytically in the weak field limit, $\cale_0/\omega^2\ll b$, or low-frequency field $\gamma\ll 1$, while in the general case the numerical solution is required. Furthermore, solutions do not exist for all values of the final momentum and arbitrary parameters $\mu$ and $\gamma$. Indeed, the second equation of (\ref{speq-3}) shows that the final momentum is determined by the time instant $\phi_0$ when the particle is released from under the barrier, and by its initial velocity $v_0$. The width of the momentum space available is determined by the field momentum, $p_{\rm max,min}\approx p_0\pm p_F$, so that when $p_0-p_F<0$ not all $\phi_0$ are solutions. Another limitation comes from the fact that Eqs.~(B.3)--(B.6) assume that the instant kinetic energy at the exit is below the instant barrier height, $v_0^2<2(U_0-\cale_0b\cos\phi_0)$; this restricts the value of the field,
$\cale_0b<\kappa_0^2/2$.

In the  rectangular barrier limit $\Delta\to 0$, the domain II vanishes and the action is determined by trajectories (\ref{trajectory1}) and (\ref{trajectory2}) and the Lagrange function,
\begin{equation}
{\cal L}=\dot x^2/2+\cale_0x\cos\varphi-U_0\Theta(b-x)\, ,
\end{equation}
where $\Theta (x)$ is the Heaviside function. Using this Lagrangian and the above determined trajectories, the classical action has to be evaluated along Eq.~(\ref{W}).

%%%%%%%%%%%%%%%%%%%%%%%%%%%%%%%%%%%%%%%%%%%%%%%%%%%%%%%%
%%%%%%%%%%%%%%%%%%%%%%%%%%%%%%%%%%%%%%%%%%%%%%%%%%%%%%%%
\section*{References}
%%%%%%%%%%%%%%%%%%%%%%%%%%%%%%%%%%%%%%%%%%%%%%%%%%%%5555

\end{document}